\RequirePackage{fix-cm}

\documentclass[twocolumn,epjc3]{svjour3}
\smartqed  
\RequirePackage{graphicx}
\RequirePackage{latexsym,amsmath,amssymb,amsbsy,graphicx,graphics,dsfont,mathrsfs}
\RequirePackage[numbers,sort&compress]{natbib}

\newcommand{\ii}{\mathrm{i}}

\newcommand{\Tr}{\mathrm{Tr}}
\journalname{Eur. Phys. J. C}
\begin{document}

\title{Resonance enhancement of neutrino oscillations due to transition magnetic moments
}

\author{A.V. Chukhnova \thanksref{e1,addr1}
        \and
        A.E. Lobanov \thanksref{e2,addr1}
}

\thankstext{e1}{e-mail: av.chukhnova@physics.msu.ru}
\thankstext{e2}{e-mail: lobanov@phys.msu.ru}

\institute{Department of Theoretical Physics, Faculty of Physics,
	Moscow State University, 119991 Moscow, Russia \label{addr1}}

\date{Received: date / Accepted: date}

\maketitle

\begin{abstract}
We prove that a resonance enhancement of neutrino oscillations in magnetic field is possible due to transition magnetic moments and demonstrate that this resonance is strictly connected to the neutrino polarization. To study the main properties of this resonance, we obtain the probabilities of transitions between neutrino states with definite flavor and helicity in inhomogeneous electromagnetic field in the adiabatic approximation.
Since the resonance is present only when the adiabaticity condition is fulfilled, we also obtain and discuss this condition.

	
\end{abstract}

The neutrino oscillations are usually described using the phenomenological theory based on the ideas by B. Pontecorvo (see, e.g. \cite{Bilenky1978}).
Within this approach the interaction with matter can be taken into account with the help of an effective potential \cite{Wolfenstein1978}, which modifies the neutrino dispersion law.  As a result, the resonance behavior of flavor oscillations, i.e. the Mikheev--Smirnov--Wolfenstein (MSW) effect \cite{MS_en}, can be observed \cite{Bethe}.
However, since neutrino is a massive particle, it is necessary to take into account not only the neutrino flavor oscillations, but also the spin rotation effect to construct a complete description of neutrino evolution.

When neutrino propagates in vacuum, in matter at rest or parallel to the direction of magnetic field \cite{AK1988}, the helicity does not change. In this case it is possible to consider the evolution of a neutrino state with a definite helicity. Nevertheless, the presence of electromagnetic field \cite{Fujikawa1980,Shrock1982}, moving or polarized matter \cite{Lobanov2001} (see also \cite{Studenikin2004}) in general case can modify neutrino spin orientation.  Spin rotation of neutrino in electromagnetic field was widely discussed about 40 years ago (see, e.g., \cite{OVV1986,Voloshin1986}). In particular, spin rotation was considered as a possible explanation of the solar neutrino problem. For the first time the external medium as a factor, which results in an actual spin precession of the neutrino, was considered in \cite{Lobanov2001}.  However, the neutrino propagation in moving matter has been studied for a long time.
The dispersion law for neutrino in moving matter was considered in \cite{Pal1989,Nieves1989_1}. In \cite{orsesmo1986_en,Semikoz1987_en,Semikoz1989_en} the polarization of the background matter was taken into account using the concept of induced neutrino magnetic moment. The value of the induced magnetic moment should be calculated for any preset composition of the background medium. In particular, in \cite{Nunokawa1997} it was calculated for the medium composed of electrons only.

Since there are correlations between flavor oscillations and spin rotation, while describing neutrino evolution in the general case it is necessary to take into account these processes simultaneously. The results of such description may be significant for astrophysics
\cite{Akhmedov1988,Volpe2015,Kartavtsev2015,Dobrynina2016,Vlasenko2014,Ternov2016,Kurashvili2017,Ternov2019}.

A rigorous quantum field theoretical description of Dirac neutrino propagation taking into account both flavor oscillations and spin rotation can be constructed within the Standard Model modification \cite{tmf2017_en,lobanov2019}, where both the mass states and their arbitrary superpositions are considered as different quantum states of an $SU(3)$ neutrino multiplet. This description can be generalized for the case of neutrino propagating in dense matter \cite{izvu2016_en,VMU2017_en} and electromagnetic field \cite{PR2020}. Then the neutrino evolution can be studied on the base of neutrino wave equation, which has the same meaning as the Dirac--Schwinger equation of quantum electrodynamics (see, e.g. \cite{B_SH_en}).  As a result of this approach in \cite{PR2020} we obtain that the cosine of the effective mixing angle for neutrino, moving in electromagnetic field with constant characteristics, changes its sign as a function of the magnetic induction, and therefore the resonance behavior of probabilities due to neutrino transition magnetic moments is possible \cite{PR2020}.
As it is well-known, the resonance can be observed when the background characteristics vary slowly in space.
In the present paper we obtain explicit formulas for neutrino spin-flavor transition probabilities in the case of inhomogeneous electromagnetic field within the adiabatic approximation, which is usually used to describe the MSW effect for neutrino in matter. We study the properties of the new resonance taking into account the restrictions on the magnetic induction, for which the adiabatic approximation is valid.

The wave equation for neutrino in electromagnetic field, which takes into account the direct interaction of neutrino multiplet with the field due to the anomalous magnetic moments and the transition magnetic and electric moments, is as follows \cite{PR2020}
\begin{multline}\label{l16}
\bigg(\ii\gamma^{\mu}{\partial}_{\mu}{\mathds I}  - {\mathds{M}} - \frac{\ii}{2}\mu_{0}F^{\mu \nu
}\sigma _{\mu \nu }{\mathds{M}}
- \frac{\ii}{2}F^{\mu \nu
}\sigma _{\mu \nu }{\mathds{M}}_{h} \\
- \frac{\ii}{2}{}^{\star\!\!}F^{\mu\nu}\sigma _{\mu \nu }{\mathds{M}}_{ah}\!\bigg)\,{\varPsi}(x) = 0.
\end{multline}
\noindent The wave functions $\varPsi(x)$ describe the $SU(3)$ neutrino multiplet as a whole. Here $\mathds{I}$ is the identity matrix, $\mathds{M}$ is the neutrino mass matrix,
${}^{\star\!}F^{\mu\nu} = -\frac{1}{2}e^{\mu\nu\rho\lambda}F_{\rho\lambda}$ is the dual tensor to the electromagnetic field tensor $F^{\mu\nu}$. The interaction with the electromagnetic field via transition magnetic and transition electric moments is taken into account by introducing the Hermitian matrix of transition magnetic moments ${\mathds{M}}_{h}$ and the matrix of the transition electric moments ${\mathds{M}}_{ah}$. In the first approximation the diagonal magnetic moments are proportional to the neutrino masses, i.e. the matrix of the diagonal magnetic moments is defined as $\mu_0 \mathds{M}$, where
\begin{eqnarray}
\mu_0 = \frac{3 e \mathrm{G_F}}{8\sqrt{2}\pi^2}.
\end{eqnarray}

The evolution of ultra-relativistic neutrinos can be studied in the quasi-classical approximation. In this case we can assume that the neutrino 4-velocity $u^\mu$, which is proportional to the neutrino kinetic momentum, is constant \cite{arlomur}. Then the evolution equation takes the form
\begin{equation}\label{t3}
\ii\dot{\varPsi} (\tau)= {\mathcal{F}}\varPsi (\tau),
\end{equation}
\noindent where
\begin{multline}\label{xl17}
{\mathcal{F}}={\mathds{M}}
-\mu_{0}\mathds{M}
\gamma^{5}\gamma^{\mu}{\,}^{\star\!\!}F_{\mu\nu}u^{\nu} \\
-\mathds{M}_{h}
\gamma^{5}\gamma^{\mu}{\,}^{\star\!\!}F_{\mu\nu}u^{\nu}
+\mathds{M}_{ah}
\gamma^{5}\gamma^{\mu}F_{\mu\nu}u^{\nu}.
\end{multline}
\noindent The dot denotes differentiation with respect to the
proper time $\tau$.

For simplicity we will consider the two-flavor model.
In the two-flavor model the mass matrix $\mathds{M}$ and the matrices of transition moments $\mathds{M}_{h}$, $\mathds{M}_{ah}$ are $2\times 2$ matrices and may be expressed in the terms of the Pauli matrices. The corresponding wave functions  $\varPsi(\tau)$ are $8$-component objects. In the mass representation
\begin{equation} \label{M_fl} \displaystyle
\begin{array}{l} \displaystyle
\mathds{M} = \frac{1}{2}(\sigma_0(m_1+m_2) - \sigma_3(m_2 - m_1)), \\ [6pt] \displaystyle
\mathds{M}_{h} = \frac{1}{2}(m_1+m_2)\mu_{1}\sigma_1, \\[6pt] \displaystyle
\mathds{M}_{ah} = \frac{1}{2}{(m_1-m_2)}\varepsilon_{1} \sigma_2,
\end{array}
\end{equation}
\noindent where $\sigma_i, i=1,2,3$ are the Pauli matrices, $\sigma_0$ is the identity $2\times2$ matrix.
For the Standard Model neutrinos the values of the coefficients $\mu_1$ and $\varepsilon_1$, which characterize the neutrino transition magnetic and electric moments, can be found in \cite{Shrock1982} (see also \cite{Giunti2015}). Within the Standard Model the transition moments are suppressed in comparison to the diagonal ones due to GIM-mechanism \cite{GIM_1970}, and the neutrino transition electric moments are smaller than the transition magnetic moments.
In this work we assume that the transition electric moments are small, and so we neglect them. In this case we are able to obtain an analytical solution of the evolution equation in the two flavor model. Possible effects of the transition electric moment are discussed further in the paper.

Then matrix \eqref{xl17} in the mass representation looks like
\begin{multline}\label{xxl17}
{\mathcal{F}}\rightarrow \frac{1}{2}\Big\{(\sigma_0(m_1+m_2) \\
 - \sigma_3(m_2 - m_1))(1-\mu_{0}
\gamma^{5}\gamma^{\mu}{\,}^{\star\!\!}F_{\mu\nu}u^{\nu}) \\
-\sigma_1 (m_1+m_2) \mu_{1}
\gamma^{5}\gamma^{\mu}{\,}^{\star\!\!}F_{\mu\nu}u^{\nu}\Big\}.
\end{multline}

 When the direction of the magnetic induction is constant, the operator
\begin{equation}\label{xxl171}
{\cal S}=\gamma^{5}\gamma^{\mu}{\,}^{\star\!\!}F_{\mu\nu}u^{\nu}/N,\quad N=\displaystyle \sqrt{\displaystyle u_{\mu}{\,}^{\star\!\!}F^{\mu\alpha}{\,}^{\star\!\!}F_{\alpha\nu}u^{\nu}}.
\end{equation}
\noindent  is an integral of motion for the evolution equation \eqref{xxl17}.  Note that for neutrino propagating orthogonally to the purely magnetic field with the induction $\bf{B}$ we have $N=u_0 |\bf{B}|$, and for neutrino propagating parallel to the magnetic field $N=|{\bf B}|$, i.e. $N$ is the absolute value of the magnetic induction in the neutrino rest frame. The operator ${\cal S}$ defines the projection of the spin on the direction of the magnetic field in the neutrino rest frame.  The corresponding neutrino polarization vector is defined as follows
\begin{equation}
\bar{s}^\mu = - {\,}^{\star\!\!}F_{\mu\nu}u^{\nu}/N.
\end{equation}
\noindent If the magnetic field varies slowly, the operator ${\cal S}$ can be considered as an approximate integral of motion, when the energy of the neutrino interaction with the field due to the magnetic moment is much larger than the inverse characteristic time of the field variation. To put it in a more formal way, the following conditions must be satisfied \cite{Lobanov2006}
\vspace{-6pt}
\begin{equation}\label{beta}
\frac{\kappa}{2\mu_0 N}\ll 1,
\end{equation}
\begin{equation}\label{alpha}
\frac{\varkappa}{\kappa}\ll 1,
\end{equation}
\noindent where 
\begin{equation}
\begin{array}{l}\displaystyle
\kappa=\sqrt{H^2 \dot{H}^2 - (H\dot{H})^2} /{N^2}, \\ \displaystyle
 \varkappa=\frac{N}{H^2 \dot{H}^2 - (H\dot{H})^2} e^{\alpha \beta \gamma \delta} \ddot{H}_{\alpha} \dot{H}_{\beta} H_{\gamma} u_{\delta}.
 \end{array}
\end{equation}
\noindent Here $H^\mu= {}^{\star\!}F^{\mu\nu} u_\nu$.

However, the neutrino Lorentz factor is rather large $u_0\gg 1$. Under the Lorentz transformations to the neutrino rest frame the longitudinal component of the magnetic field remains the same, while the orthogonal component increases proportionally to $u_0$. Hence, the direction of the magnetic induction in the neutrino rest frame is almost orthogonal to the velocity of this reference frame, except for the case of the neutrino propagating precisely in the direction of the magnetic field or against it. Except for this case, the directions of the first and second derivatives of the magnetic induction in the neutrino rest frame are also almost orthogonal to the direction of the neutrino velocity, and condition \eqref{alpha} is satisfied.

In paper \cite{PR2020} we obtain solutions of the evolution equation in the case of constant background conditions with the help of the resolvent $U(\tau)$
\vspace{-6pt}
\begin{equation}\label{sf1}
\varPsi(\tau)=\frac{1}{\sqrt{2 u_0}}\, U(\tau) \varPsi_{0}.
\end{equation}
\noindent \vskip-6pt Here $\varPsi_{0}$ is a constant object, which defines the neutrino initial state and in the two flavor model has $8$ components. For a neutrino pure state with a definite initial polarization it can be presented in the form
\begin{equation}\label{sf2}
\begin{array}{l}\displaystyle
\varPsi_0=\frac{1}{2}(1-\gamma^{5}\gamma_{\mu}{s}_{0}^{\mu}
) (\gamma_\mu u^\mu+1) \left(\psi^0\otimes e_{j}\right), \\ [6pt]
 \bar{\varPsi}_0\varPsi_0= 2.
 \end{array}
\end{equation}
\noindent \vskip-6pt Here $\psi^0$ is a constant bispinor,
$e_{j}$ is an arbitrary unit vector in the two-dimensional vector space over the field of complex numbers, and ${s}_{0}^{\mu}$ is a $4$-vector of neutrino polarization such that $(u{s}_{0})=0$.

When the external conditions vary slowly, it is also possible to write the resolvent $U(\tau)$. Since the operator $\cal{S}$ with the eigenvalues $\zeta$ can be considered as an integral of motion, we can study the evolution of neutrino states with definite $\zeta$ independently. Note that the states with definite $\zeta$ in the general case are not the states with definite helicity.

The matrix, which determines the evolution equation for the states with definite $\zeta$, can be diagonalized in the mass representation at a given point $\tau$ using the matrix $\mathds{U}^{(eff)}_\zeta(\tau)$
\begin{equation}
\mathds{U}^{(eff)}_\zeta (\tau) = \left(
\begin{matrix} \cos{\theta'}_\zeta(\tau) & \sin{\theta'}_\zeta (\tau) \\ - \sin{\theta'}_\zeta (\tau)
& \cos{\theta'}_\zeta (\tau) \end{matrix}
\right).
\end{equation}
\noindent Here ${\theta}'_\zeta (\tau)$ is an effective mixing angle in electromagnetic field in the mass representation for the states with definite $\zeta$  (for more detail see \cite{PR2020}). The value of the angle ${\theta}'_\zeta (\tau)$ is determined by the relations
\begin{equation}\label{cond}
\begin{array}{l}\displaystyle
X'_{\zeta}(\tau) =\sin{2\theta'_\zeta(\tau)}, \\ [6pt]
Y'_{\zeta}(\tau)=\cos{2\theta'_\zeta(\tau)},
\end{array}
\end{equation}
\noindent where
\begin{equation}\label{XYZ}
\begin{array}{l}
Y'_{\zeta}(\tau)=\displaystyle\frac{1}{Z_{\zeta}(\tau)}\Big (\big(m_{2}-m_{1}\big) \big( 1-\zeta\mu_{0}N(\tau) \big) \Big), \\ [8pt]
X'_{\zeta}(\tau)=\displaystyle\frac{1}{Z_{\zeta}(\tau)}\Big( -\zeta\mu_{1}N(\tau) \big(m_{2}+m_{1}\big)\Big), \\ [8pt]
Z_{\zeta}(\tau)=\Big\{\Big (\big(m_{2}-m_{1}\big) \big( 1-\zeta\mu_{0}N(\tau) \big) \Big)^{2} \\
\phantom{Z_{\zeta}(\tau)======} +\Big(\big(m_{2}+m_{1}\big)\mu_{1}N(\tau)\Big)^{2}\Big\}^{1/2},
\end{array}
\end{equation}
\noindent Then for $\sin{\theta'_\zeta(\tau)}$ and $\cos{\theta'_\zeta(\tau)}$ we have
\begin{equation}
\begin{array}{l}\displaystyle
\sin{\theta'_\zeta(\tau)} = \mathrm{sgn}X'_\zeta(\tau) \sqrt{(1-Y'_\zeta(\tau))/2}, \\ [6pt] \displaystyle
 \cos{\theta'_\zeta(\tau)} = \sqrt{(1+Y'_\zeta(\tau))/2}.
 \end{array}
\end{equation}

The adiabatic approximation is valid when $	|\dot{\theta}_\zeta| \ll Z_\zeta$. We can introduce the adiabaticity parameter $\Gamma$ for neutrino in electromagnetic field by analogy with what is usually done for neutrino in matter (see, e.g., \cite{Giunti2007}). The adiabaticity condition takes the form
\begin{equation}\label{usl_ad}
\Gamma = \frac{2 Z_\zeta^3}{\mu_1 |\dot{N}(\tau)| (m_2^2-m_1^2)} \gg 1.
\end{equation}
If the external conditions vary slowly, then the Hamiltonian of the system is almost diagonal at every point.  Then to write down the solution with rather high accuracy, it is enough to diagonalize the matrix of the equation at the current point $\tau$ and to perform the inverse transformation at the initial point $\tau=0$ \cite{Fedoryuk_en}. The method is absolutely similar to what is usually done to describe MSW resonance. Using this approach, we can obtain the resolvent $U'(\tau)$ in the mass representation in the adiabatic approximation.

The matrices in \eqref{M_fl}, \eqref{xxl17} and the resolvent $U(\tau)$ are presented in the mass representation. To obtain the same matrices in the flavor representation one should use the transformation
\begin{equation}\label{UU}
U(\tau) = \mathds{U} U'(\tau) \mathds{U}^\dag.
\end{equation}
\noindent Here $\mathds{U}$ is the Pontecorvo--Maki--Nakagawa--Sakata mixing matrix, which in the two-flavor model is defined by the vacuum mixing angle $\theta$ as follows
\begin{equation}\label{U}
\mathds{U} = \left( \begin{matrix} \cos{\theta} & \sin{\theta} \\
- \sin{\theta} & \cos{\theta} \end{matrix} \right),
\end{equation}
\noindent
To obtain the resolvent in the flavor representation one needs to take into account relation \eqref{UU}. So, in adiabatical approximation we derive the following expression for the resolvent of the wave equation, which describes the neutrino multiplet in inhomogeneous magnetic field
\begin{multline}\label{res_fl}
U(\tau) = \sum\limits_{\zeta=\pm 1} e^{-\ii \int\limits_{0}^{\tau}   (T_{\zeta}(\tilde{\tau})/2) d\tilde{\tau} }
\times \Big(C_\zeta \cos(\theta_\zeta(\tau)-\theta_\zeta(0)) \\
 - \ii S_\zeta \sigma_1 \sin(\theta_\zeta(\tau)+\theta_\zeta(0))
	+\ii C_\zeta \sigma_2 \sin(\theta_\zeta(\tau)-\theta_\zeta(0))
	\\+ \ii S_\zeta \sigma_3 \cos(\theta_\zeta(\tau)+\theta_\zeta(0)) \Big) \Lambda_\zeta.
\end{multline}
\noindent The values of the mixing angles in the flavor representation are given by the relation
\begin{equation}\label{theta}
\theta_\zeta(\tau) = \theta + \theta'_\zeta(\tau).
\end{equation}
\noindent In Eq. \eqref{res_fl} the following notations are used
\begin{equation}
\begin{array}{l}\displaystyle
S_\zeta = \sin\!\!{\int\limits_{0}^{\tau}\!\! {\big(Z_\zeta(\tilde{\tau})}/{2}\big)\, d \tilde{\tau}},
 C_\zeta = \cos\!\!{\int\limits_{0}^{\tau} \!\!{\big(Z_\zeta(\tilde{\tau})}/2\big)\, d \tilde{\tau}},\ \\[12pt] \displaystyle
T_{\zeta}(\tau)=\displaystyle\big(m_{2}+m_{1}\big) \big( 1-\zeta\mu_{0}N(\tau)\big),
\end{array}
\end{equation}
\noindent  which generalize the notations used in paper \cite{PR2020} for constant electromagnetic field.
The projection operators on the eigenstates of operator ${\cal S}$ are defined by the expressions
\begin{equation}
\Lambda_\zeta = \frac{1}{2}\left(1 + \zeta {\cal S} \right), \quad [\gamma^\mu u_\mu,\Lambda_\zeta]=0, \quad\zeta=\pm 1.
\end{equation}

To calculate the probabilities of transitions between states with definite flavor and polarization, we use quasi-classical spin-flavor density matrices introduced similarly to the quasi-classical spin density matrices
(see \cite{Lobanov2006})
\begin{equation}\label{rho4}
\rho_{\alpha}(\tau)=\frac{1}{4u^{0}}U(\tau)\big(\gamma^{\mu}u_{\mu}+1\big)\left(1-
\gamma^{5}\gamma_\mu {s}^{\mu}_{0}
\right){\mathds P}_{0}^{(\alpha)}\bar{U}(\tau)
\end{equation}
\noindent In this formula ${s}^{\mu}_{0}$ defines the initial polarization state of the neutrino, the projection operator ${\mathds P}_{0}^{(\alpha)}$ defines its initial flavor state and the resolvent $U(\tau)$ is given by Eq. \eqref{res_fl}.
The probability of a transition from the state $\alpha$ to the state $\beta$ in the time $\tau$ is determined by the following relation
\begin{equation}\label{ver}
W_{\alpha \rightarrow\beta}=\Tr\left\{\rho_{\alpha}(\tau)\rho^{\dag}_{\beta}
(\tau=0)\right\}.
\end{equation}
\noindent Note that since in our model these states of the neutrino multiplet are pure states, all the final formulas may also be obtained using the neutrino multiplet wave functions.

In the flavor representation the projection operators on  states with the definite  flavor take the form
\begin{equation}\label{M_fl2}
\begin{array}{l} \displaystyle
\mathds{P}^{(\alpha)}_{0} =   \frac{1}{2}(1 +\xi_{\alpha}\sigma_3),  \displaystyle
 \mathds{P}^{(\beta)}_{0} =   \frac{1}{2}(1 +\xi_{\beta}\sigma_3), \\ [6pt] \displaystyle
  \xi_{\alpha},\xi_{\beta} =\pm 1.
\end{array}
\end{equation}
\noindent To obtain the projection operators on the initial and final state with the electron flavor one should choose $\xi_\alpha, \xi_\beta=1$, otherwise $\xi_\alpha, \xi_\beta=-1$. We  assume that in the initial and final states neutrino has a definite helicity, i.e.
\begin{equation}\label{t61}
\begin{array}{l}\displaystyle
{s}_{0}^{(\alpha)\mu}=\zeta_{\alpha}{s}^{\mu}_{sp},\quad {s}_{0}^{(\beta)\mu}=\zeta_{\beta}{s}^{\mu}_{sp}, \\[4pt] \displaystyle
{s}^{\mu}_{sp}=
\{|{\bf u}|,u^{0}{\bf u}/|{\bf u}|\}, \quad \zeta_{\alpha}, \zeta_{\beta}=\pm 1,
\end{array}
\end{equation}
\noindent where the values $\zeta_\alpha,\zeta_\beta=1$ correspond to the right-handed neutrino  and $\zeta_\alpha,\zeta_\beta=-1$ correspond to the left-handed neutrino. The spin-flavor transition probabilities can be presented in the form
\begin{multline}\label{l31}
W_{\alpha \rightarrow\beta}=\frac{1+\xi_\alpha\xi_\beta}{2}\frac{1+\zeta_\alpha\zeta_\beta}
{2}W_1 + \frac{1+\xi_\alpha\xi_\beta}{2}\frac{1-\zeta_\alpha\zeta_\beta}{2}W_2 \\ \!\!\!
+ \frac{1-\xi_\alpha\xi_\beta}{2}\frac{1+\zeta_\alpha\zeta_\beta}{2}W_3 + \frac{1-\xi_\alpha\xi_\beta}{2}\frac{1-\zeta_\alpha\zeta_\beta}{2}W_4,
\end{multline}
\noindent where
\begin{equation}\label{028}
\begin{array}{l}\displaystyle
W_1=\frac{1}{8}\Big((1\!-\!\zeta_\alpha(\bar{s}s_{sp}))^2\big(1+
C_{+1}^{2}\cos{2\Delta_{+1}(\tau)}  \\ [4pt] \displaystyle
\phantom{W_1}+
S_{+1}^{2}\!\cos{2\Theta_{+1}(\tau)}\big)\\ [4pt] \displaystyle
\phantom{W_1}\displaystyle \!+\!(1\!+\!\zeta_\alpha(\bar{s}s_{sp}))^2\big(1+C_{-1}^{2}\!\cos{2\Delta_{-1}(\tau)} \\ [4pt] \displaystyle
\phantom{W_1}\!+\!
S_{-1}^{2}\!\cos{2\Theta_{-1}(\tau)}\big) \Big) \\ [4pt] \displaystyle
\phantom{W_1}\displaystyle\!+\!\frac{1}{2}(1\!-\!(\bar{s}s_{sp})^2)\big(\xi_\alpha F_1(\tau)\sin{\Phi(\tau)}+D_1(\tau)\cos{\Phi(\tau)} \big),\\[8pt]
\displaystyle
W_2\!=\!\frac{1}{8}\Big( (1\!-\!(\bar{s}s_{sp})^2)\big(1+C_{+1}^{2}\cos{2\Delta_{+1}(\tau)} \\ [4pt] \displaystyle
\phantom{W_2}+
S_{+1}^{2}\cos{2\Theta_{+1}(\tau)}\big)\\ [4pt]
\phantom{W_2}\displaystyle + (1\!-\!(\bar{s}s_{sp})^2)\big(1+C_{-1}^{2}\cos{2\Delta_{-1}(\tau)} \\ [4pt] \displaystyle
\phantom{W_2}
+S_{-1}^{2}\!\cos{2\Theta_{-1}(\tau)}\big) \Big) \\  \displaystyle
\phantom{W_4}-\!\frac{1}{2}(1\!-\!(\bar{s}s_{sp})^2)\big(\xi_\alpha F_1(\tau)\sin{\Phi(\tau)} + D_1(\tau)\cos{\Phi(\tau)} \big), \\[8pt]
\displaystyle
W_3\!=\!\frac{1}{8}\Big( (1\!-\!\zeta_\alpha(\bar{s}s_{sp}))^2\big(1-C_{+1}^{2}\!\cos{2\Delta_{+1}(\tau)} \\ [4pt] \displaystyle
\phantom{W_3}\!-\!
S_{+1}^{2}\!\cos{2\Theta_{+1}(\tau)}\big)\\ [4pt] \displaystyle
\phantom{W_3}\displaystyle\!+\!(1\!+\!\zeta_\alpha(\bar{s}s_{sp}))^2
\big(1-C_{-1}^{2}\!\cos{2\Delta_{-1}(\tau)} \\ [4pt] \displaystyle
\phantom{W_3} -
S_{-1}^{2}\!\cos{2\Theta_{-1}(\tau)}\big) \Big)  \\ \displaystyle
\phantom{W_3}+\!\frac{1}{2}(1\!-\!(\bar{s}s_{sp})^2)\big(\xi_\alpha F_2(\tau)\sin{\Phi(\tau)} \!+\!D_2(\tau)\cos{\Phi(\tau)}\big), \\[8pt]
\displaystyle
W_4\!=\!\frac{1}{8}\Big((1\!-\!(\bar{s}s_{sp})^2)\big(1-C_{+1}^{2}\!\cos{2\Delta_{+1}(\tau)} \\ [4pt] \displaystyle
\phantom{W_4} -
S_{+1}^{2}\!\cos{2\Theta_{+1}(\tau)}\big)\\ [4pt] \displaystyle
\phantom{W_4}\displaystyle\!+\!(1\!-\!(\bar{s}s_{sp})^2)\big(1-C_{-1}^{2}\!\cos{2\Delta_{-1}(\tau)} \\ [4pt] \displaystyle
\phantom{W_4}\!-\!
S_{-1}^{2}\!\cos{2\Theta_{-1}(\tau)}\big) \Big) \\ [4pt] \displaystyle
\phantom{W_4}\!-\!\frac{1}{2}(1\!-\!(\bar{s}s_{sp}))^2)\big( \xi_\alpha F_2(\tau) \sin{\Phi(\tau)} \!+\!D_2(\tau)\cos{\Phi(\tau)} \big).
\end{array}
\end{equation}
\noindent Here we use the notations
\begin{equation}\label{2z}
\begin{array}{l}\displaystyle
\Phi(\tau)=\mu_0 (m_1+ m_2)\int\limits_{0}^{\tau} N(\tilde{\tau}) d \tilde{\tau},\\ [4pt]
F_1(\tau)= C_{+1}S_{-1}\cos\Delta_{+ 1}(\tau)\cos\Theta_{-1}(\tau) \\
\phantom{F_1(\tau)}
- S_{+1}C_{-1}\cos\Theta_{+1}(\tau)\cos\Delta_{-1}(\tau), \\

D_1 (\tau)= C_{+1}C_{-1} \cos\Delta_{+1}(\tau)\cos\Delta_{-1}(\tau) \\
\phantom{D_1(\tau)}\,
+
S_{+1}S_{-1}\cos\Theta_{+1}(\tau)\cos\Theta_{-1}(\tau), \\
F_2(\tau) = C_{+1}S_{-1}\sin\Delta_{+1}(\tau)\sin\Theta_{-1}(\tau)\\
\phantom{F_2(\tau)}\,
-S_{+1}C_{-1}\sin\Theta_{+1}(\tau)\sin\Delta_{-1}(\tau), \\
D_2(\tau) = C_{+1}C_{-1}\sin\Delta_{+1}(\tau)\sin\Delta_{-1}(\tau) \\
\phantom{D_2(\tau)}\,
+ S_{+1}S_{-1}\sin\Theta_{+1}(\tau)\sin\Theta_{-1}(\tau),
\end{array}
\end{equation}
\noindent where
\begin{equation}
\Delta_{\pm 1}(\tau) = \theta_{\pm 1}(\tau)-\theta_{\pm 1}(0), \quad
\Theta_{\pm 1}(\tau)= \theta_{\pm 1}(\tau)+\theta_{\pm 1}(0).
\end{equation}

 In the general case the formulas are rather complicated even in the two-flavor model. The spin-flavor transition probabilities can be simplified for a neutrino, which is generated in the region with high values of magnetic induction and detected in vacuum. Then the final value of the mixing angle in the flavor representation is  $\theta_{\zeta}({\tau})=\theta$, and the initial value we denote as $\theta_{\zeta}^{0}=\theta_{\zeta}(0)$. As we usually have no information concerning the exact location of the neutrino generation process, we should average the probabilities
over the proper time $\tau$. Thus, the averaged values of the probabilities are
\begin{equation}\label{32}
\begin{array}{l} \displaystyle
W_1 = \frac{1}{8}\Big((1-\zeta_\alpha(\bar{s}s_{sp}))^2 \big(1+ \cos{2\theta}\, \cos{2\theta}_{+1}^{0}\big) \\ [4pt] \displaystyle \phantom{W_1 =}+  (1+\zeta_\alpha(\bar{s}s_{sp}))^2\big(1+ \cos{2\theta}\, \cos{2\theta}_{-1}^{0}\big)\Big),\\ [4pt]\displaystyle
W_2 = \frac{1}{8}(1- (\bar{s}s_{sp})^2)\big(2 + \cos{2\theta} (\cos{2\theta}_{+1}^{0} +  \cos{2\theta}_{-1}^{0}) \big),\\ [6pt] \displaystyle
W_3 = \frac{1}{8}\Big((1-\zeta_\alpha(\bar{s}s_{sp}))^2 \big(1- \cos{2\theta}\, \cos{2\theta}_{+1}^{0}\big) \\ [4pt] \displaystyle
\phantom{W_3=} + (1+\zeta_\alpha(\bar{s}s_{sp}))^2\big(1- \cos{2\theta}\, \cos{2\theta}_{-1}^{0}\big)\Big),\\[4pt] \displaystyle
W_4 = \frac{1}{8}(1- (\bar{s}s_{sp})^2)\big(2 -\cos{2\theta} (\cos{2\theta}_{+1}^{0} +  \cos{2\theta}_{-1}^{0} ) \big).
\end{array}
\end{equation}

For high energy neutrinos the assumption $(\bar{s}s_{sp})=0$ is valid with high accuracy, except for a narrow region of angles when neutrino velocity and the vector of magnetic induction
are almost parallel (see \cite{PR2020}).
Therefore, the probabilities take the form
\begin{equation}\label{ave1}
\begin{array}{l}\displaystyle
W_{1,2} = \frac{1}{8}\left(2+ \cos{2\theta}(\cos{2\theta}_{+1}^{0} +\cos{2\theta}_{-1}^{0})\right), \\[6pt] \displaystyle
W_{3,4} =  \frac{1}{8}\left(2- \cos{2\theta}(\cos{2\theta}_{+1}^{0} +\cos{2\theta}_{-1}^{0}) \right).
\end{array}
\end{equation}
\noindent Taking into account Eq.\eqref{cond} and Eq.\eqref{theta}, we have
\begin{equation}
\cos 2\theta_{\zeta}^0= Y'_\zeta  \cos{2\theta}-X'_\zeta \sin{2\theta}.
\end{equation}
\noindent In the explicit form \eqref{theta} can be presented as follows
\begin{multline}\label{expltheta}
\cos 2\theta_{\zeta}^0 = \Big( (m_{2}-m_{1}) ( 1-\zeta\mu_{0}N(0) )\cos{2\theta}\\
+\zeta(m_{2}+m_{1})\mu_{1}N(0) \sin{2\theta} \Big) \\
\times
\Big(\!\big((m_{2}-m_{1})(1-\zeta\mu_{0}N(0)) \big)^{2}
	\! \\
	+\big((m_{2}+m_{1})\mu_{1}N(0)\big)^{2}\! \Big)^{\!-1/2}\!\!.
\end{multline}
\noindent Note that when the neutrino state can be described as a superposition of the mass eigenstates, the effective mixing angles $\theta_\zeta$ are equal to their vacuum values.

The theoretical predictions for the Standard Model neutrinos give $\mu_1/\mu_0 \sim 10^{-4}$ (see \cite{Shrock1982}), and so it can be expected that
\begin{equation}\label{rel}
r=\frac{\mu_1 (m_1+m_2)}{\mu_0 (m_2-m_1)}\ll 1.
\end{equation}
\noindent In this case in the first approximation the transition probabilities take the form
\begin{equation}\label{W12SM}
\begin{array}{l}
W_{1,2}\!=\!\displaystyle\frac{1}{8}\!\bigg(2\!+\!\Big(1\!+\!\mathrm{sgn}\big(1\!-\!\frac{\!\mu_0 (m_2\!-\!m_1)\!}{m_2-m_1} N(0)\big)\Big)\!\cos^2{2\theta\!}\bigg),\\ [10pt]
W_{3,4}\!=\!\displaystyle\frac{1}{8}\!\bigg(2\!-\!\Big(1\!+\!\mathrm{sgn}
\big(1\!-\!\displaystyle\frac{\!\mu_0 (m_2\!-\!m_1)\!}{m_2-m_1} N(0)\big)\Big)\!\cos^2{2\theta\!}\bigg).	
\end{array}
\end{equation}
\noindent Obviously, when $\mu_0 N (0) \approx 1$, expression \eqref{W12SM} is not valid, and the exact formula \eqref{ave1} is necessary.

However, if the adiabaticity condition is not fulfilled, even for initial fields, which exceed the resonance value, the resonance behavior of transition probabilities will not be observed. Since in the resonance region $Z_\zeta \approx \mu_1 N (m_1+m_2)$, the adiabaticity parameter $\Gamma$ in \eqref{usl_ad} becomes proportional to $\mu_1^2$.
		Hence, though $\mu_1$ is absent in Eq. \eqref{W12SM}, if we put $\mu_1=0$ from the start, the adiabaticity condition can not be fulfilled in the resonance region and we will not obtain the resonance behavior of the transition probabilities.
	
The coefficient $r$ defined by Eq. \eqref{rel} actually determines the region of the fields, where $\mu_1$ can not be neglected. That means, the greater is the parameter $r$, the wider is the resonance region. Note that for smaller values of $r$ the adiabaticity condition \eqref{usl_ad} results in stronger restrictions on the possible value of the field gradient.

\begin{figure*}
	\begin{minipage}{0.48\textwidth}
		\vspace*{-11pt}	\includegraphics[width=\textwidth]{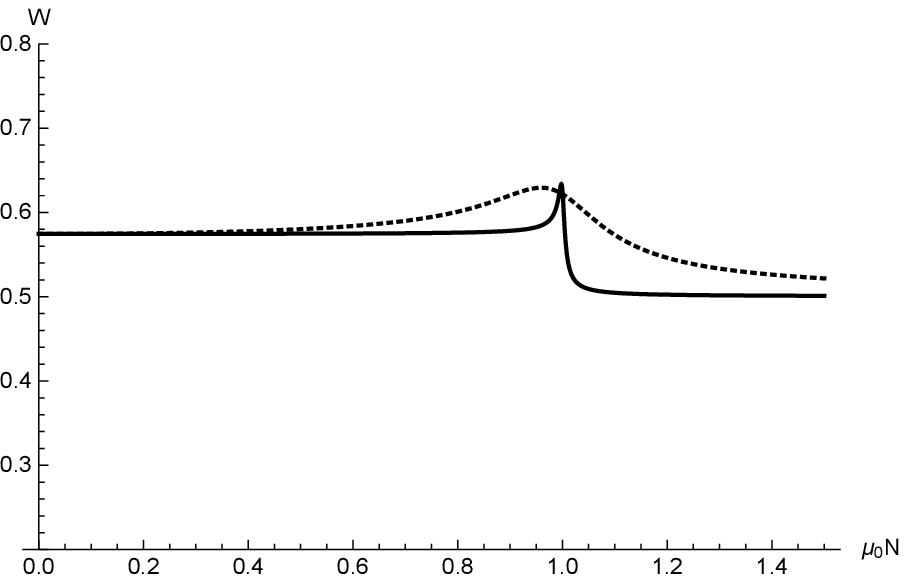}	
		\caption{The dependence of the flavor survival probability on the initial magnetic induction for a neutrino propagating orthogonally to the magnetic field for $r=0.005$ (the solid line) and for $r=0.1$ (the dashed line).}\label{Fig1}
	\end{minipage} \hskip 15 pt
	\begin{minipage}{0.48\textwidth}
	\vspace*{-2pt}	\includegraphics[width=\textwidth]{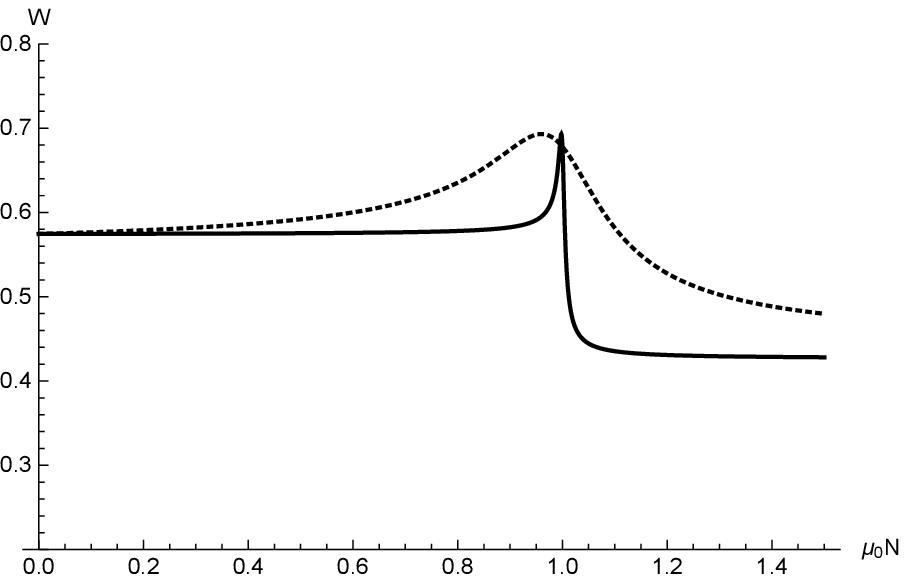}
		\caption{
			The dependence of the flavor survival probability on the initial magnetic induction for a left-handed neutrino propagating against the direction of the magnetic field (or a right-handed neutrino propagating in the direction of the field) for $r=0.005$ (the solid line) and for $r=0.1$ (the dashed line).
		}\label{Fig2}
	\end{minipage}
\end{figure*}

 In Fig. \ref{Fig1} the behavior of the neutrino flavor-survival probability $W_1+W_2$ for different values of $r$ is demonstrated in the case when the neutrino velocity is orthogonal to the magnetic induction vector.
 Here we choose the value of the vacuum mixing angle such that $\sin^2 \theta = 0.307$.

In the present paper we neglect the transition electric moment, since no explicit analytical solution can be found for arbitrary values of the transition electric moment. However, the resonance behavior of the flavor transition probabilities  will still be present, since it is determined by the diagonal elements of the matrices in the wave equation expressed in the mass representation. For the Standard Model neutrinos the effect of the transition magnetic and electric moments becomes significant only for the values of the initial magnetic field, which are in the resonance region $\mu_0 N \sim 1$. That is, the resonance denominator of the effective mixing angle contains the value $Z_\zeta$, which can be approximated as
	\begin{multline}
	Z_\zeta \approx \Big(\big( (m_2-m_1)(1-\zeta \mu_0 N)\big)^2+ \big((m_1+m_2)\mu_1 N\big)^2  \\
 + \big(\alpha (m_2-m_1)\varepsilon_1 \tilde{N}\big)^2\Big)^{1/2},
	\end{multline}
	\noindent where $|\alpha|\leq 1$. Indeed, the interaction with the electric transition moment is determined by the spin operator
	\begin{equation}
	\tilde{\cal S}=-\gamma^{5}\gamma^{\mu}F_{\mu\nu}u^{\nu}/\tilde{N}, \quad \tilde{N}=\displaystyle \sqrt{\displaystyle u_{\mu}F^{\mu\alpha}F_{\alpha\nu}u^{\nu}}.
	\end{equation} The square of this operator, as well as the square of any spin operator, is equal to unity, and therefore this approximation can be obtained. Obviously, the electric moment will not make the resonance disappear.  When the transition electric moment is taken into account, no spin integral of motion exists for Eq. \eqref{xl17}. Thus, the neutrino can not propagate in any spin eigenstate. This may lead to the following consequences. Firstly, the value of the resonance field may be slightly changed. Secondly, the probabilities $W_1$ and $W_2$ may differ from each other, as well as $W_3$ may differ from $W_4$ (see \eqref{ave1}).

For high values of magnetic field  all the spin-flavor transition probabilities become equal
\begin{equation}\label{1/4}
W_1 = W_2 = W_3 = W_4= \frac{1}{4}.
\end{equation}
\noindent This means that  all the information about neutrino initial state is lost.
Expression \eqref{1/4} follows directly from Eq. \eqref{ave1} not only when condition \eqref{rel} is satisfied.

Although for most situations the condition $(\bar{s} s_{sp}) =0$ is satisfied, there is a region of angles, for which neutrino velocity and the vector of magnetic induction
are almost parallel and this condition is not fulfilled. As it is already mentioned, for ultra-relativistic neutrinos this region of angles is very narrow, and even a small angular deviation  makes neutrino behave almost like in the case of orthogonal propagation \cite{PR2020}. Because of this instability, from the phenomenological point of view the situation is hardly of any practical value.

 However, understanding this case is very important since it helps us to explain, why in the general case the number of the neutrinos of the second flavor does not predominate the number of the neutrinos of initial flavor even for the magnetic fields higher than the resonance field (see Fig. \ref{Fig1}). Let us consider the limiting case, when the neutrino  moves  either in the direction of the magnetic field or against it.
In these cases the helicity operator becomes an integral of motion, and the spin-flip transitions are absent since $(\bar{s}s_{sp})= \pm 1$, i.e. $W_2=W_4=0$.

For neutrino moving against the direction of the magnetic field we have $(\bar{s}s_{sp})= 1$. Hence,
\begin{equation}\label{inthedir}
\begin{array}{l}\displaystyle
W_1 =\frac{1-\zeta_\alpha}{4}(1+ \cos{2\theta}  \cos{2\theta}_{+1}^{0})   \\ [6pt] \displaystyle
\phantom{W_1 =}+ \frac{1+\zeta_\alpha}{4}(1+ \cos{2\theta}  \cos{2\theta}_{-1}^{0}) ,\\[6 pt] \displaystyle
W_3 = \frac{1-\zeta_\alpha}{4}(1 -\cos{2\theta}  \cos{2\theta}_{+1}^{0})   \\ [6pt] \displaystyle
\phantom{W_3 =} + \frac{1+\zeta_\alpha}{4} (1 -\cos{2\theta}  \cos{2\theta}_{-1}^{0}) .
\end{array}
\end{equation}
\noindent If inequality \eqref{rel} holds, we can write the approximate expressions for the transition probabilities
\begin{equation}\label{inthedir00}
\begin{array}{l}\displaystyle
W_1 = \frac{1\!-\!\zeta_\alpha}{4}\bigg(1\!+\!\mathrm{sgn}\Big(1\!-\!\frac{\!\mu_0 (m_2\!-\!m_1)\!}{m_2-m_1} N(0)\Big) \cos^{2}{\!2\theta}\bigg) \\ \displaystyle
\phantom{W_1 =} +  \frac{1\!+\!\zeta_\alpha}{4}(1\!+\!\cos^{2}{2\theta}) ,\\ [6pt] \displaystyle
W_3 = \frac{1\!-\!\zeta_\alpha}{4}\bigg(1\!-\!\mathrm{sgn}\Big(1\!-\!\frac{\!\mu_0 (m_2\!-\!m_1)\!}{m_2-m_1} N(0)\Big) \cos^{2}{\!2\theta}\bigg) \\ \displaystyle
\phantom{W_3 =} + \frac{1\!+\!\zeta_\alpha}{4}(1\!-\!\cos^{2}{2\theta}).
\end{array}
\end{equation}
\noindent  According to these formulas, the resonance is present for left-handed neutrinos, while for right-handed neutrinos it is absent.

For neutrino moving in the direction of the magnetic field we have $(\bar{s}s_{sp})= -1$. The probabilities in this case can be obtained if we change $\zeta_{\alpha}\rightarrow - \zeta_\alpha$  in \eqref{inthedir}, \eqref{inthedir00}. So the resonance is present only for right-handed neutrinos, which are not observed in experiments. The dependence of the flavor-survival probability $W_1+W_2$ on the value of the magnetic induction is demonstrated in \mbox{Fig. \ref{Fig2}}.

Therefore, an important conclusion can be derived. The presence of the resonance depends on the neutrino polarization. Let us look back at formula \eqref{ave1}. For a left-handed particle propagating orthogonally to the magnetic field the probabilities to observe the spin projection in the direction of the field and opposite to it are both equal to $1/2$, and the neutrino spin-flavor transition probabilities are the sum of the resonant and non-resonant terms. It is a well-known fact that the MSW resonance is observed for the left-handed neutrinos only. Since for the neutrinos in matter at rest the helicity is conserved, the problem of the correlation between the possibility of the resonance behavior and the actual polarization of the particle did not arise in the studies of the  resonance in matter. In this sense the description of resonance in the magnetic field is a more complicated problem, than of the MSW resonance for neutrino propagation in matter at rest. In this paper we consider in detail the case of neutrinos with definite initial and final helicities. However, our approach enables one to study the states with any polarization. It can be seen from \eqref{32} that the transition probabilities exhibit full resonance when the initial neutrino spin is directed along the magnetic induction vector, and the resonance is absent when the neutrino spin is directed opposite the magnetic induction no matter what the direction of neutrino velocity is.
This may be important for the case of low neutrino energies, when the neutrino chiral states differ significantly from the neutrino helicity states.

The resonance condition is $\mu_0 N \approx 1$, and therefore it is determined by the value of the magnetic induction in the neutrino rest frame. Since the value of $\mu_0$ given by the Standard Model is very small, extremely high values of the magnetic field are required. The estimates of the values of neutrino energy and the magnetic field, which are necessary for the resonance to take place, are discussed in detail in \cite{PR2020}. For neutrino propagating orthogonally to the magnetic field the
	resonance is reached when $u_0 B/B_0 \approx 1.3 \times 10^{13}$, where $B_0 = 4.41 \times 10^{13}$ Gauss is the Schwinger magnetic field. That is, $B\approx 5.8\times 10^{26} (m_\nu/\mathcal{E}_\nu)$ Gauss, where $m_\nu$ is the average neutrino mass and $\mathcal{E}_\nu$ is the neutrino energy. For neutrino with the mass $m_\nu=0.033$ eV in the magnetic field $B = 10^{16}$ Gauss the energy about $1.9$ GeV is needed for the resonance to take place. These estimates indicate that not only the resonance, but even the spin oscillations are very unlikely to be observed in currently known magnetars, since the characteristic length of spin oscillations seems to be much greater than the size of the corresponding astrophysical objects. It is also very important that since for the Standard Model neutrinos $r\ll 1$, the adiabaticity condition \eqref{usl_ad}
\begin{equation}\label{usl_ad_SM}
\frac{r^{2}(m_2^2 -m_1^2)}{{\cal E}_\nu}  \gg \frac{|dN/dx|}{N}.
\end{equation}
\noindent seems to give a very strict restriction on the value of the field gradient $dN/dx$. Here ${\cal E}_\nu$ is the neutrino energy, and $x$ is the spacial coordinate along the neutrino trajectory.
Even when the magnetic field and the neutrino energy are high enough for the resonance to take place, if relation \eqref{usl_ad_SM} is violated, no resonance will be present.
The resonance might become observable for Standard Model neutrinos if some exotic compact objects with higher values of magnetic fields or larger size than the currently known magnetars are discovered. In our opinion taking into account this resonance as well as neutrino spin rotation might also be interesting in the studies of the Early Universe, since these effects change the flavor and helicity characteristics of the neutrino flux and therefore might influence the particle composition of the background matter.

However, there are models of New Physics, which predict greater values of neutrino magnetic moments than the Standard Model. Since the Standard Model theoretical prediction is $\displaystyle \mu_\nu= \mu_0 m_\nu \sim 3 \cdot 10^{-19} \left(\frac{m_\nu}{1 \text{eV}} \right) \mu_B$ and the current experimental restriction on the neutrino magnetic moment is $\mu_\nu < 2.9 \cdot 10^{-11} \mu_B$ \cite{Agostini2017, Beda2013}, these models can not be excluded.  If such New Physics exists, then near magnetars the spin-flip effect and even the resonance behavior of neutrino transition probabilities may be observed (see Fig. \ref{Fig1}). For such models the mass matrix and the matrix of the diagonal magnetic moments may be not proportional. Obviously, our results are valid within the models of New Physics with the values of the transition moments, which are significantly smaller than the diagonal magnetic moments. For this case our results are also applicable. To obtain the expressions for transition probabilities within such models one should only replace in all our formulas
\begin{equation}
\begin{array}{l}
\mu_0 m_1 \rightarrow \mu_{11}, \qquad  \mu_0 m_2 \rightarrow \mu_{22}, \\ [6pt]
 \mu_1 (m_1+m_2) \rightarrow 2 \mu_{12}.
\end{array}
\end{equation}
\noindent Here $\mu_{11}$, $\mu_{22}$ are the diagonal magnetic moments, and $\mu_{12}$ is the transition magnetic moment. The resonance condition in this case takes the form
\begin{equation}
N\frac{|\mu_{22}- \mu_{11}|}{|{m_2-m_1}|}=|1 - r\tan{2\theta}|^{-1},
\end{equation}
\noindent where $r=2\mu_{12}/|\mu_{22}- \mu_{11}|$. For neutrinos propagating parallel to the magnetic field this condition was obtained in \cite{AK1988}.
If  $r \gtrsim 1$  the adiabaticity condition \eqref{usl_ad} takes the form
\begin{equation}\label{usl_ad_nonSM}
\frac{(m_2^2 -m_1^2)}{{\cal E}_\nu}  \gg \frac{|dN/dx|}{N}
\end{equation}
\noindent and the following effects might become observable.
The spin rotation effect might become significant for neutrinos produced inside the magnetars or other compact objects with the high values of the magnetic field. Therefore, the total flux of observable neutrinos of all flavors might become less than the initial flux, since only left-handed neutrinos interact with a terrestrial detector. For solar neutrinos the possibility of this effect was discussed in \cite{OVV1986}. The flavor composition of the flux of the neutrinos produced inside some compact object might differ significantly from the the same flux measured in terrestrial conditions due to the resonance studied above. However, what effects will really be observed depends on the definite value of $\mu_\nu$.
It should be emphasized that in the models of New Physics with large neutrino transition moments for the high values of the magnetic field all the averaged spin-flavor transition probabilities become approximately equal to $1/4$ similar to the case of Standard Model neutrinos (see \eqref{1/4}).

In this paper we generalize the approach used in \cite{PR2020}, where the  problem was studied in the case of constant external conditions. We find analytical expressions for  solutions  of the neutrino wave equation in magnetic field in adiabatic approximation in two-flavor model taking into account transition magnetic moments.  We derive the formulas for the transition probabilities and indeed obtain a resonance enhancement of neutrino oscillations due to  transition magnetic moments, which was predicted in paper \cite{PR2020}.
We show that the type of the resonance is determined by the neutrino polarization and in the case of extra-high value of magnetic fields the averaged values of all spin-flavor transition probabilities are equal to $1/4$.

\begin{acknowledgements}
The authors are grateful to A.V. Bori\-sov, E.M. Murchikova, I.P. Volobuev, and V.Ch. Zhukovsky for fruitful discussions. A.V.C. acknowledges support from the Foundation for the advancement of theoretical physics and mathematics ``BASIS'' (Grant No. 19-2-6-100-1).
\end{acknowledgements}

\end{document}